\documentclass[a4paper]{jpconf}

\usepackage[utf8]{inputenc}
\usepackage{mathrsfs}
\usepackage{bm}
\usepackage{graphicx}
\usepackage{subfigure}
\usepackage[english]{babel}

\begin{document}

\title{Probing the hue of the stochastic magnetization dynamics}

\author{Stam Nicolis$^1$, Julien Tranchida$^{1,2}$ and Pascal Thibaudeau$^{2}$}

\address{$^1$CNRS-Laboratoire de Mathématiques et Physique Théorique (UMR 7350), Fédération de Recherche "Denis Poisson" (FR2964), Département de Physique, Université de Tours, Parc de Grandmont, F-37200, Tours, FRANCE}
\address{$^2$CEA DAM/Le Ripault, BP 16, F-37260, Monts, FRANCE}

\ead{stam.nicolis@lmpt.univ-tours.fr; julien.tranchida@cea.fr; pascal.thibaudeau@cea.fr}

\begin{abstract}
The Fokker--Planck equation describes the evolution of a probability distribution towards equilibrium--the flow parameter is the equilibration time. Assuming the distribution remains normalizable for all times, it is equivalent to an open hierarchy of equations for the moments. Ways of closing this hierarchy have been proposed; ways of explicitly solving the hierarchy equations have received much less attention. In this paper we show that much insight can be gained by mapping the Fokker--Planck equation to a Schrödinger equation, where Planck's constant is identified with the diffusion coefficient.\\\\
\end{abstract}

\section{Introduction\label{intro}}
Consider a stochastic process, described by a differential equation on a one-dimensional variable $x$, driven by an additive Gaussian noise $\eta(t)$ of intensity $D$,
\begin{equation}
\label{Langevin}
\frac{dx}{dt}=-\frac{dV(x)}{dx}+\eta(t).
\end{equation} 
This induces a probability measure, $P(x,t)$, on the space of configurations,$\{x\}$, that satisfies a master equation:
\begin{equation}
\label{mastereq}
\frac{\partial P(x,t)}{\partial t} = -\frac{\partial J(x,t)}{\partial x}
\end{equation}
where $J(x,t)$ is the corresponding probability current. In this equation, both $P(x,t)$ and $J(x,t)$ depend on the potential $V(x)$. Under certain assumptions it's possible to show that the current, $J$, is a local function of the density $P$, in particular and it is given by the expression
\begin{equation}
\label{FP}
J(x,t)=-D\frac{\partial P(x,t)}{\partial x} - \frac{dV(x)}{dx}P(x,t),
\end{equation}
in which case the master equation is called the ``Fokker--Planck equation''\cite{risken_fokker-planck_1989}.

This equation describes the approach to an  equilibrium $P_\mathrm{eq}(x)$ with $t\ge 0$, provided $V(x)$ satisfies the condition
\begin{equation}
\label{PartitionFun}
Z=\int_{-\infty}^{\infty}\,dx\,e^{-\frac{V(x)}{D}} < \infty
\end{equation}
in which case the equilibrium distribution is, indeed, given by the expression 
\begin{equation}
\label{Peq}
P_\mathrm{eq}(x)=\frac{1}{Z}e^{-\frac{V(x)}{D}}
\end{equation}
Now, we shall work in units, where the coefficient $D$ in front of $\partial^2 P/\partial x^2$ is set to 1.

The boundary conditions, for this to hold is that $\lim_{x\to\pm\infty}\,P(x,t)=0$, i.e. that $V(x)$ confines ``sufficiently strongly''.  For different boundary conditions the equilibrium distribution can, of course, be completely different, or, even, not exist at all. 

In many practical situations, $P(x,t)$ and $P_\mathrm{eq}(x)$ can be characterized by their moments, $\langle x^m\rangle(t)$ and 
$\langle x^m\rangle_\mathrm{eq}$ respectively. Additionally, the moments are, typically, obtained, directly, by measurements, in either real or computer experiments and the probability densities are not known beforehand, but rather their properties are to be deduced from the properties of the moments. In such situations it is useful to write the evolution equations for the moments of $P(x,t)$ with the following recursion relation:
\begin{equation}
\label{hierarchyFP}
\begin{array}{l}
\displaystyle
\frac{d}{dt}\left\langle x^m\right\rangle(t)\equiv\int\,dx\,x^m\,\frac{\partial P(x,t)}{\partial t}=
-m\int\,dx\,x^{m-1}\left(\frac{\partial P(x,t)}{\partial x} + \frac{dV(x)}{dx}P(x,t)\right)=\\
\displaystyle
\hskip2.2truecm
m(m-1)\left\langle x^{m-2}\right\rangle  - m\left\langle x^{m-1}\frac{dV(x)}{dx}\right\rangle
\end{array}
\end{equation}
for $m=2,3,\ldots$. For $m=1$ one must be a bit more careful and go back to the definition of $\langle x\rangle(t)$ itself to obtain 
\begin{equation}
\label{FirstMom}
\frac{d}{dt}\left\langle x\right\rangle = -\left\langle\frac{\partial V}{\partial x}\right\rangle.
\end{equation}
The equations~(\ref{hierarchyFP}) and (\ref{FirstMom}) provide an infinite set of ordinary differential equations for the moments $\langle x^m\rangle(t)$. In these equations, also, boundary terms have been assumed not to contribute and must be studied separately, once provided an infinite set of initial conditions $\langle x^m\rangle(0)\equiv u_m$. 

It is possible to generalize this calculation to higher dimensions: Suffice to interpret the terms in the Fokker--Planck equation appropriately, at first, for flat manifolds, by replacing derivatives by gradients and products by scalar products:
\begin{equation}
\label{multidim}
\begin{array}{l}
\displaystyle \frac{\partial^2P}{\partial x^2}\to\Delta P(\bm{x},t)\\
\displaystyle \frac{\partial}{\partial x}\left(\frac{\partial V}{\partial x}P\right)\to \nabla\cdot\left(P(\bm{x},t)\nabla V(\bm{x})\right)
\end{array}
\end{equation}
 For more general manifolds one simply has to introduce the appropriate metric and all expressions can be written in covariant form. In this case the moments become tensors, e.g.
\begin{equation}
\label{multimoms}
\left\langle x_{I_1}^{m_1}x_{I_2}^{m_2}\cdots x_{I_k}^{m_k}\right\rangle(t), 
\end{equation}
and are subject to coupled differential equations as well.

For $V(\bm{x})$ a polynomial, whatever the number of dimensions, equations (\ref{hierarchyFP}) with proper initial conditions, are linear equations in the moments and describe their flow towards their equilibrium values. The existence of a normalizable equilibrium distribution implies that the equilibrium values should be independent of the initial conditions, provided the moments do exist\footnote{The normalizability of the density is a necessary, but not sufficient, condition for the existence of the moments.}.

For $V(\bm{x})$ not a polynomial, the equations are non--linear in the moments. However it's not this property that makes them hard to solve; the polynomial case is hard to solve, also. 

The real problem for solving eqs.~(\ref{hierarchyFP}) and (\ref{FirstMom}), resp. eq.~(\ref{multimoms}),  is that the equations are coupled and the matrix of the linear system is not of finite rank. Only two cases allow the hierarchy to be solved explicitly: the case $V(x)=0$ and the case where $V(x)=kx^2/2$, with $k>0$, because this second case can be reduced to the first one, by a change of variables. 
In fact these two cases illustrate the nature of the problem: each moment decays at a different time scale, set by its order.  It is this property that provides the real obstruction to ``closing the hierarchy''. And the reason is the absence of ``mixing'', that depends on the interplay between diffusion and the interactions described by the potential. In the two cases at hand, this is due to the fact that the potential can, in fact, be eliminated, which eliminates mixing.

This was noted many years ago~\cite{nicolis_closing_1998} and it was pointed out that the interplay between the mixing properties of the diffusion term of Fokker--Planck and the ergodic properties of  the potential, can make certain moments, of different orders, tend to decay at the same rate. This is the theoretical foundation of the attempts for ``closing'' the hierarchy and thus reducing the system of moment equations to finite rank. In recent work~\cite{tranchida_closing_2016} and work in progress, this approach was shown to be relevant, in practice, also, for closing the hierarchy of moments of the magnetization of a nanomagnet described by the Landau--Lifshitz--Gilbert equation. That this is possible at all in this system isn't a straightforward extension of the formalism.  

The reason is that, for this system, noise does not appear additively, like in eq.~(\ref{Langevin}), but multiplicatively; also that, instead of being white--with ultra--local two--point correlation function--it's colored, with a finite auto--correlation time. These properties mean that it is not true that the master equation  for the probability density, $P(x,t)$, is of Fokker--Planck form--the expression for the current, $J(x,t)$ must be found and the finite autocorrelation time means that it won't be a local function of $P(x,t)$. Therefore the hierarchy for the moments is the only way of computing them and thus of reconstructing $P(x,t)$. Since the Fokker--Planck equation is a partial differential equation, it is possible to solve it directly--numerically if need be--to machine precision and compute the moments directly. If the master equation is not known explicitly at all, attempting to close the hierarchy seems to be a way to reliably compute the moments. 

Another way, however, is possible: to try and parametrize density $P(x,t)$ and current $J(x,t)$ on the basis of the symmetries of the problem and deduce an equation that can be solved--if need be numerically. From its solution, the moments can then be directly obtained, thereby eliminating having to deal with the hierarchy altogether. As a bonus, such a scheme would provide a way of describing the approach to synchronization of the moments involved. 

In this contribution we shall show that a practical way of computing the moments of the master equation is by associating it to a Schrödinger equation, where the role of Planck's constant is played by the diffusion coefficient. 
 
\section{Effective map}

If the master equation is of Fokker--Planck form, and $P_\mathrm{eq}(x)\equiv\exp(-V(x))$ is normalizable,  
it is possible to perform a change of variables and rewrite the Fokker--Planck equation as a Schrödinger equation in imaginary time~\cite{PhysRev.150.1079}
\begin{equation}
\label{FPSchroed}
P(x,t)\equiv e^{-V(x)/2}\phi(x,t)\Rightarrow \frac{\partial\phi}{\partial t} = \frac{\partial^2\phi}{\partial x^2} + U(x)\phi
\end{equation}
with boundary conditions $\lim_{|x|\to\infty}\,\phi(x,t)=0$ and the potential $U(x)$ is given by the expression 
\begin{equation}
\label{FPpotential}
U(x)=-\frac{1}{4}\left(\frac{\partial V}{\partial x}\right)^2+\frac{1}{2}\frac{\partial^2 V}{\partial x^2}
\end{equation}
We deduce that  $\langle x^m\rangle(t)$ is given by the explicit expressions
\begin{equation}
\label{newmoms}
\left\langle x^m\right\rangle(t)=\frac{\int_{-\infty}^\infty\,dx\,e^{-V(x)/2}\phi(x,t)x^m}{\int_{-\infty}^\infty\,dx\,e^{-V(x)/2}\phi(x,t)} \Rightarrow
\left\langle x^m\right\rangle_\mathrm{eq}=\frac{\int_{-\infty}^\infty\,dx\,e^{-V(x)/2}\phi_0(x)x^m}{\int_{-\infty}^\infty\,dx\,e^{-V(x)/2}\phi_0(x)} 
\end{equation}
where $\phi_0(x)=\exp(-V(x)/2)$ is the ground state wave-function. This is the solution to the problem, provided that $\phi_0(x)$, indeed exists, i.e. that the fluctuations do not affect the potential. Similarly, for higher dimensional target spaces, the  equal time moments are given by the expressions
\begin{equation}
\langle x_{I_1}^{m_1}\cdots x_{I_k}^{m_k}\rangle(t)=\frac{\int d\bm{x}\,e^{-V(\bm{x})/2}\phi(\bm{x},t)x_{I_1}^{m_1}\cdots x_{I_k}^{m_k}}{\int d\bm{x}\,e^{-V(\bm{x})/2}\phi(\bm{x},t)}.
\end{equation}
These expressions can then be readily evaluated, thereby obviating the technical need for closing the hierarchy.  What the framework of closing the hierarchy provides is the motivation to understand which groups of moments synchronize towards equilibrium.

For the case of a nanomagnet in a magnetic field, however, what the appropriate expression for the master equation should be is not obvious at all, since the force term does not derive from a potential. 
In previous work on the subject we have shown~\cite{tranchida_closing_2016,tranchida_colored-noise_2016,thibaudeau_non-markovian_2016} that it is possible to derive the hierarchy of equations for the moments directly from the Landau--Lifshitz--Gilbert equation and, using the techniques of ref.~\cite{nicolis_closing_1998}, to close the hierarchy and solve the equations numerically. Obtaining the master equation, however, is considerably harder, since it is not a local equation for the probability density, even for vanishing autocorrelation time. Therefore we need to understand what are the physical properties that such a master equation must have and attempt to parametrize the space of its solutions in another way, beyond Fokker--Planck.

For the case of a vector potential, as in the Landau--Lifshitz--Gilbert equation
\begin{equation}
\label{LLG}
\frac{d\bm{x}}{dt} = \bm{\omega}\times\bm{x} - \lambda\left(\bm{\omega}\times\bm{x}\right)\times\bm{x}
\end{equation}
things become more complicated, because the RHS cannot be written as the gradient of a potential. Furthermore, the noise term doesn't appear explicitly. So this doesn't seem to be a stochastic equation at all. Stochastic effects could appear, were this system chaotic~\cite{perez2015effect}; this isn't the case, however, since, for fixed precession frequency, the component of the magnetic moment parallel to the frequency vector is conserved and the transverse components define an integrable system. 

If the precession frequency can fluctuate, however, then stochastic effects can appear and the master equation must be found by appealing to the symmetries of the problem. The key symmetry is gauge invariance. 

The precession frequency $\bm{\omega}\equiv\bm{\omega}(t)$, can, in fact, be  identified with the magnetic field, that can be expressed in terms of a vector potential $\bm{A}(t)$, such as $\bm{\omega}(t)\equiv\nabla\times\bm{A}(t)$. Gauge invariance is the symmetry $\delta_\Lambda\bm{A}=\nabla\Lambda$. Stochastic effects are described by assuming that $\bm{\omega}(t)\equiv\bm{\omega}_0+\bm{\Omega}(t)$, with $\bm{\Omega}(t)$ drawn from a Gaussian process with color defined by a finite auto--correlation time, 
\begin{equation}
\label{colored_noise}
\left\langle\Omega_I(t)\Omega_J(t')\right\rangle = \frac{D}{\tau}\delta_{IJ}\exp\left(-\frac{|t-t'|}{\tau}\right)
\end{equation}
This disorder induces a probability measure on the space of configurations for the magnetization, $P(\bm{x},t)$, that satisfies a master equation--that is not a Fokker--Planck equation for finite autocorrelation time. We would like to explore the possibility of deducing it as the continuity equation of an appropriate Schrödinger equation. This time there is not any scalar potential, however.  But it is known how to write the Schrödinger equation for a particle in a magnetic field, in terms of its vector potential, including the anisotropy relevant for the Zeeman effect: one introduces the covariant derivatives and the anisotropy involves the coupling to angular momentum. The mixing effects are captured, once more, by the diffusion coefficient, while the effects of the finite autocorrelation time are captured by averaging the moments over the distribution of the vector potential. 

The advantage of this formulation is that it allows us to describe what happens, when the vector potential, $\bm{A}(\bm{x},t)$ is drawn from some distribution. This distribution must be invariant under gauge transformations, therefore, depend only on the precession field, derived from the vector potential. For a Gaussian distribution this means that it is, necessarily, of the form
\begin{equation}
\label{rho_of_A}
\rho(\bm{A}) = \exp\left(-\frac{1}{2}\int\,dt\,dt'\,\left(\bm{\omega}[\bm{A}](t)-\bm{\omega}_0\right)G(t-t')\cdot\left(\bm{\omega}[\bm{A}](t')-\bm{\omega}_0\right)\right)
\end{equation}
with $\omega_I[\bm{A}]=\varepsilon_{IJK}\nabla_J A_K$ and $G(t-t')$ the functional inverse of the 2--point function $C(t-t')$ of the precession field:
\begin{equation}
\label{2ptFunctionField}
\left\langle \left(\omega_I(t)-\omega_{0I}\right)\left(\omega_J(t')-\omega_{0J}\right)\right\rangle =\delta_{IJ}C(t-t')
\end{equation}
with the other correlation functions defined by Wick's theorem,
if we assume that the precession field fluctuates about the value $\langle\bm{\omega}[\bm{A}](t)\rangle=\bm{\omega}_0$.

Since the probability density for the precession field is Gaussian, it is, in principle, possible to integrate out the vector potential and obtain an effective action for the magnetic moment. If the noise is ``colored'', then the effective action will, indeed,  be non-local and the equation satisfied by the probability density, in the saddle point approximation will not be of Fokker--Planck type, for this reason.

In this way we obtain the effective action for the averaged wavefunction $\psi(\bm{x},t)$, from which we can compute the density, $P(\bm{x},t)$, whose equal time moments can then be computed. This is clearly a formulation consistent with all symmetries of the problem. While analytical expressions are difficult to obtain, numerical techniques, that are reliable to machine precision, are known. 

\begin{figure}[thp]
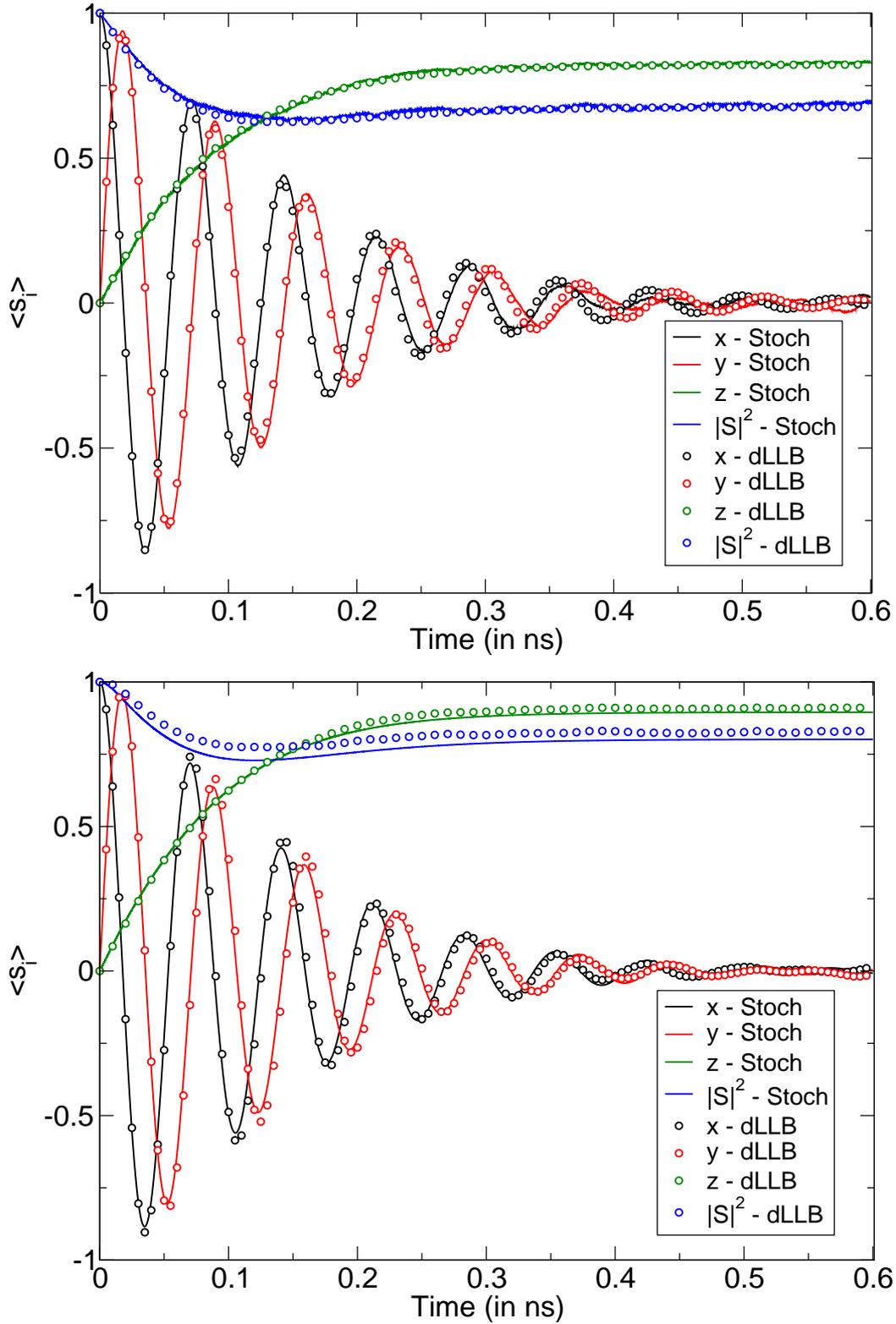

\subfigure{\includegraphics[scale=0.5]{Fig1.eps}}
\subfigure{\includegraphics[scale=0.5]{Fig2.eps}}
\caption[]{ $\langle\bm{s}\rangle(t)$ for weakly colored noise, $\tau=10^{-7}$ ns (upper panel) vs.  strongly colored noise, $\tau=10^{-2}$ ns (lower panel). Taken from~\cite{tranchida_colored-noise_2016}. }
\label{colored_noisevevs}
\end{figure}

The details of this approach will be presented in future work. It is possible to establish a benchmark for it, by studying the approach to equilibrium of the equal--time moments, for ``small'' , as well as ``large'' values of $\tau$. The technical details can be found in ref.~\cite{tranchida_colored-noise_2016}; they are illustrated in fig.~\ref{colored_noisevevs} where $s_i$ stands for the i-th component of the average magnetization (i.e. the first moment).

\section{Conclusions\label{conclusions}}
We have described several, {\em a priori}, different ways for computing the equal time moments of the magnetization, that determine its probability distribution, when stochastic perturbations must be described with a finite auto--correlation time. On the one hand we have implemented a numerical method, based on an efficient integration of the Langevin equation and averaging over the realizations of the noise; on the other hand, we have implemented a method for testing a theoretical approach, based on chaotic dynamical systems and have found that the two approaches, while providing complementary information, are in agreement; and the framework put in place allows to improve the approximations in a systematic way. We have, also, set up a theoretical framework,  based on a Schr\"odinger equation, that exploits the gauge invariance of the system and where the diffusion coefficient plays the role of Planck's constant. While, in quantum mechanics, this approach is, of course, well known, the fact that, from a mathematical viewpoint, it is of much more general validity, hasn't been, yet, appreciated. 

The finite auto--correlation time of the noise implies, in general, that the probability current is a non--local functional of the density, which is natural in the Schr\"odinger approach.  This leads to  new numerical and conceptual tools for studying non--Markovian dynamics (cf. also~\cite{Roma:2014ek}), beyond the vicinity of the white--noise limit~\cite{fox_functional-calculus_1986} and can lead to new insights for further developing the functional integral approach~\cite{Aron:2014uo}) and will be presented more fully in future work.

\section*{References}
\bibliographystyle{iopart-num}
\bibliography{ClosureMoments}
\end{document}